\documentstyle[12pt]{article}

\begin{document}

\begin{center}
{\Large {\bf
The Fubini-Furlan-Rossetti sum rule}}
\\
{\Large {\bf
reexamined}}
\end{center}
\vskip 0.20in
\begin{center}
Richard A. Arndt and Ron L.\ Workman \\
Department of Physics  \\
Virginia Polytechnic Institute and State University \\
Blacksburg, Virginia 24061-0435 \\
\end{center}
\vskip 0.20in
\begin{abstract}
We review the status of a Fubini-Furlan-Rossetti sum rule for the
photoproduction of pions from nucleons,
indicating how well this sum rule is satisfied using a recent
analysis of pion photoproduction data. Our results are discussed in light
of a recent paper comparing the Gerasimov-Drell-Hearn and spin-dependent
polarizability sum rules.
\end{abstract}
\vfil
\eject

The development of Chiral Perturbation theory (ChPT) and extended current
algebra has led to a renewed interest in a number of sum rules derived
in the 1960's. Examples include the Gerasimov-Drell-Hearn\cite{gdh} (GDH)
and Weinberg\cite{wsr} sum rules, as well as sum rules for the nucleon
electric, magnetic and spin-dependent polarizabilities\cite{sand}.
In the following, we will consider
another sum rule\cite{ffr}, due to Fubini, Furlan and Rossetti (FFR),
which has not attracted as much attention. We will also briefly comment on a
recent paper\cite{sand} which compares the GDH sum rule with a
related one for the nucleon spin-polarizability.

Our study of the FFR sum rule was preceded by a reexamination of the
GDH sum rule in the context of extended current algebras. While the
GDH sum rule was first derived from a dispersion relation (unsubtracted)
and the low-energy theorem (LET) for Compton scattering, it was later
obtained from the commutation relations of vector current densities.
In Ref.\cite{clw} the extended current algebra of
Chang and Liang\cite{cl} was shown to imply a modified GDH sum rule.
(It was observed\cite{ks} that
modified currents would lead to modified sum rules soon after the
original GDH sum rule appeared.)
An estimation of this modification was shown to account for an apparent
discrepancy\cite{wa} in the original sum rule.

In their discussion of modified sum rules, the authors of Ref.\cite{ks}
mentioned in passing that a similar procedure  could be used to determine
modifications to the FFR sum rule.
This sum rule relates nucleon magnetic moments to an integral over the
invariant amplitude ($A_1$) for single-pion photoproduction.
The FFR sum rule has the form\cite{diff}
\begin{equation}
g_A \left( { {e \kappa^{V,S}} \over {2 M} } \right) =
{ {2 f_{\pi} } \over {\pi} } \int Im \; A^{(+,0)}_1 (\nu) \; {{d\nu}\over \nu}
\end{equation}
where $\kappa^{V,S}$ is the isovector (isoscalar) anomalous
magnetic of the nucleon, given by $(\kappa_p \mp \kappa_n)/2$.
The invariant amplitude $A_1$ corresponds to the amplitude associated with
$\gamma_5 \gamma \cdot \epsilon \gamma \cdot k$
in the paper of Chew, Goldberger, Low, and Nambu\cite{cgln}.
The required isospin combinations are given\cite{cgln}, in terms of
charge-channel information, by
\begin{equation}
A_1^{(+,0)} = \left( A_1 \; (\gamma p\to \pi^0 p) \; \pm \;
                   A_1 \; (\gamma n\to \pi^0 n) \right) \; / \; 2.
\end{equation}
Here the amplitude for photoproduction of $\pi^0 n$ states is inferred from
measurements in the three other charge channels.

Empirical evaluation
of the integral in Eq.(1) is (in principle) much simpler than the
integral in the GDH sum rule $-$ which involves contributions from multi-pion
final states. Unfortunately, there are two problems which make a precise
check more difficult. Unlike the GDH sum rule, the FFR sum rule is not
exact.  It requires use of the Goldberger-Treiman relation\cite{gt}. In
addition, convergence of the associated integral is expected to be less
rapid than was found in the GDH sum rule.

Regardless of the above qualifications,
early attempts to evaluate the integral in Eq.(1) were encouraging. An
analysis\cite{ffr} using the $P_{33}(1232)$ and $D_{13}(1520)$ resonances
found good agreement for both $\kappa^V$ and $\kappa^S$.
A subsequent study\cite{ag},
using an early multipole analysis\cite{walk}, found 85\% of the prediction
for $\kappa^V$ but did not present results for the isoscalar combination.
In Ref.\cite{ag} the threshold behavior of the multipoles was modified
by a factor to account for a non-zero pion mass\cite{extrap}.

This brings us to the reason for re-examining the FFR sum rule.
If this sum rule is in fact valid, as the early studies suggest,
it puts a constraint on the single-pion photoproduction multipoles.
The integral for $\kappa^S$, in particular, involves a delicate
cancellation between amplitudes, and thus provides a useful test of the
single-pion production input to the GDH sum rule. Other tests of the
GDH integral (including the $\pi \pi N$ contributions)
have been made recently by Sandorfi et al.\cite{sand}.
In Ref.\cite{sand},
the multipole input to the GDH and spin-dependent polarizability
sum rules was compared to predictions from ChPT\cite{meis}.
The integrals in these sum rules involve the difference of helicity 3/2
and 1/2 total cross sections
weighted by different powers of the photon energy.
Though the difference of proton and neutron GDH sum rules has
a serious problem\cite{wa}, the difference of proton and neutron
spin-dependent polarizabilities was found to agree with ChPT.

While such comparisons are interesting, our poor knowledge of the $\pi \pi N$
contribution is an impediment. Early estimates of the $\pi \pi N$
contributions were based upon the resonance spectrum found in analyses of
$\pi N$ elastic scattering data. This neglects contributions from
possible `missing states' which couple very weakly to the $\pi N$ channel.
(Though the FFR sum rule is not exact, we at
least understand the approximation (PCAC) we are making.)

We first evaluated the integrals given in Eq.~(1) using the results of a
multipole analysis\cite{caps} from 1993.
The integral giving $\kappa^V$ is heavily dominated by the
$P_{33} (1232)$ contribution, while the integral corresponding to
$\kappa^S$ shows sensitivity to cancellations between the low and
high energy parts. The result for $\kappa^V$ was 1.96, remarkably
close to the predicted value. (Exact agreement is not expected.)
The integral corresponding to $\kappa^S$, however, gave
only about 25 \% of its predicted value. This discrepancy is
beyond what one would expect from the use of PCAC.

The FFR and GDH integrals were subsequently
re-evaluated using our most recent results
from the analysis of single-pion photoproduction data. This more recent
analysis was the result of a critical examination of the entire database,
and resulted in a significantly improved fit\cite{sp95}.
The analysis was also extended to 2 GeV in order to better regularize
the results at our end-point energy of 1.8 GeV.
The isovector components of the FFR and GDH integrals
were not changed significantly by the more recent analysis.
The GDH isovector-isoscalar (VS) component, however,
was qualitatively different in the region just below the $D_{13}(1520)$
resonance and also near the high energy limit (1.8 GeV) of the multipole
solution. While these changes did not alter the character of the VS result
(the disagreement of this component with the sum rule prediction remains),
the isoscalar FFR integral is very sensitive to the $D_{13}(1520)$ resonance
region and was substantially different
($\kappa^S$ changed from $-0.015$ to $-0.069$).
The energy dependence of the FFR integrand is displayed in Fig.~1.

In summary, the FFR sum rule for $\kappa^V$  appears
to be reasonably well satisfied,
as was the case for the isovector GDH sum rule.
We also see that the FFR integral does not converge as quickly as
the analogous GDH integral. The isoscalar result is less certain.
The existence of significant structure apart from the
$D_{13}$ resonance suggests that early success\cite{ffr}
with this component of the FFR sum rule was fortuitous.
While the isoscalar integral gives a result of the correct magnitude (compare
the above results with $(\kappa_p + \kappa_n)/2 \approx \; -0.06$),
we cannot claim quantitative agreement.
However, we should note that the isoscalar component of the FFR sum rule
appears to have less problems than the VS component of the GDH sum rule.
(Estimates of the VS component disagree with the sum rule in both magnitude
and sign.) This tends to weaken arguments that require a large discrepancy
in the single-pion photoproduction multipoles in order to explain the
GDH discrepancy. It would be helpful if
high-quality photoproduction measurements could be extended
a further 1~GeV in order to
test the convergence of both the FFR and GDH sum rules.

If extended current algebra does indeed contribute
to the FFR sum rule (as suggested in Ref.\cite{ks}), the results
presented here should provide a useful test for the form
proposed by Chang and Liang\cite{cl}. While the isoscalar FFR sum rule
would likely provide the most sensitive check on any such contribution,
the phenomenological evaluation of the associated integral is not yet
sufficiently stable for more than an order-of-magnitude test.

R.W. acknowledges useful discussions with Lay Nam Chang.
This work was supported in part by the U.S. Department of
Energy Grant DE-FG05-88ER40454.

\eject
\newpage
{\Large\bf Figure captions}\\
\newcounter{fig}
\begin{list}{Figure \arabic{fig}.}
{\usecounter{fig}\setlength{\rightmargin}{\leftmargin}}
\item
{Integrand for Eq.(1) using multipole solution SP95 (see text).
(a) Contribution to $\kappa^V$, (b) Contribution to $\kappa^S$.}
\end{list}

\begin{thebibliography}{99}
\bibitem{extrap} As the extrapolation $m_{\pi} \to 0$
is not well defined, we have used the physical photoproduction amplitudes.
\bibitem{caps} These results were obtained from VPI solution FA93.
This solution was also used in the study of Ref.\cite{sand}.
\bibitem{neut} Both the VS component of the GDH sum rule and
the FFR sum rule for $\kappa^S$ are sensitive to neutron target data.
It should be noted that our multipole solutions do not
agree well with the existing neutron target data above 800 MeV.
\end{thebibliography}
\end{document}